\documentclass[3p,times,twocolumn]{elsarticle}

\usepackage{ecrc}

\volume{00}

\firstpage{1}

\journalname{Nuclear and Particle Physics Proceedings}

\runauth{J.-F. Paquet}

\jid{nppp}

\jnltitlelogo{Nuclear and Particle Physics Proceedings}

\usepackage{amssymb}

\usepackage[figuresright]{rotating}

\begin{document}

\begin{frontmatter}



\dochead{}

\title{Electromagnetic probes of heavy ion collisions: recent developments}


\author{Jean-Fran\c{c}ois Paquet}

\address{Department of Physics and Astronomy, Stony Brook University, Stony Brook, New York 11794, USA}

\begin{abstract}
The current status of photon and dilepton emission in ultra-relativistic heavy ion collisions is reviewed, and recent  developments are highlighted. The importance of emissions at early, intermediate and late times is emphasized.
\end{abstract}


\end{frontmatter}


\section{Introduction}

There is compelling evidence that a plasma of deconfined nuclear matter is created in ultra-relativistic heavy ion collisions at the Relativistic Heavy Ion Collider (RHIC) and the Large Hadron Collider (LHC). Analyses of soft hadron measurements suggest that this plasma approaches local thermal equilibrium rapidly ($\tau \sim 0.1$--$1$~fm) and proceeds to a relatively long ($\tau \sim 10$~fm) phase of hydrodynamical expansion~\cite{deSouza:2015ena}. Most soft hadrons are understood to be produced toward the end of this expansion, giving them a limited sensitivity to the properties of the earlier, hotter parts of the plasma.

Electromagnetic probes, on the other hand, have a long mean-free-path compared to the size of the plasma. They can thus escape and reach the detectors with minimum interactions with the medium, giving them the potential to provide information about the earlier stages of the deconfined plasma~\cite{Gale:2009gc}. As such, they complement hadronic observables and provide additional constraints on the spacetime evolution of the plasma and its properties.

In ultra-relativistic proton-proton collisions, which are used as reference for heavy ion collisions, electromagnetic emission is generally well understood. Photons are produced as the result of hard parton interactions (``prompt photons'') and through decays of unstable hadrons. This latter source can usually be constrained experimentally and removed from measurements, producing the observable known as ``direct photons''. This subtraction is difficult and often not done for dileptons, and hadronic decays constitute a large contribution to dilepton observables. Other sources of dileptons include hard parton interactions (``Drell-Yan dileptons'') and decays of heavy quarks.

All sources of electromagnetic probes found in proton-proton collisions are also encountered in ultra-relativistic heavy ion collisions, all modified in one way or the other by the formation of the deconfined plasma and by cold nuclear matter effects. Additional sources of emission are induced by the deconfined plasma, such as the thermal photons and dileptons radiated by the plasma during its hydrodynamical expansion~\cite{Gale:2009gc}. The imprint left by the plasma on thermal emissions, prompt photons, decays of heavy quarks and hadrons, and other possible sources of radiation, all constitute additional levers to help constrain the properties of the plasma. 

In this contribution, the current status of electromagnetic probes in heavy ion collisions is reviewed. A discussion of early and late stage emissions is provided to put in context recent work on the subject~\cite{preEq}. Photons are used as example, but most considerations apply to dileptons as well. The discussion is primarily oriented toward ultra-relativistic heavy ion collisions at the LHC and at the top RHIC energy, where the description of the spacetime evolution of the plasma in terms of hydrodynamics is best understood.

\section{Electromagnetic probes: a status update}

Soft electromagnetic measurements at the RHIC (Au-Au, $\sqrt{s_{NN}}=200$~GeV) consist of the low $p_T^\gamma$ direct photon spectra and anisotropies (mainly $v_2$ and $v_3$), the dilepton invariant mass spectra, and a limited measurement of the dilepton $v_2$ at small invariant mass~\cite{BatheHP2016}.
The direct photon spectra was also measured at the LHC for Pb-Pb collisions with $\sqrt{s_{NN}}=2760$~GeV, and the direct photon $v_2$ is under active investigations~\cite{Bock:2016jus}, with some preliminary measurements having been presented in the past.

It is generally believed that direct photon observables are dominated by thermal photons at low $p_T^\gamma$ . The exponential dependence of the soft direct photon spectra is often cited as support for this conclusion, although photons produced through other mechanisms, such as interactions of hard partons with the plasma~\cite{Turbide:2007mi}, could realistically produce a similar signature. Stronger support for thermal photons as the main source of low $p_T^\gamma$ photons is the similarity of the direct photon $v_2$ and the charged hadron one in terms of size and shape: few mechanisms have been shown to be able to produce  \emph{both} a significant contribution to the low $p_T^\gamma$ photon spectra with a large $v_n$ closely resembling the charged hadron's.

As for the dilepton invariant mass spectra, in Au-Au collisions at $\sqrt{s_{NN}}=200$~GeV, there are two windows around $0.3-0.6$~GeV and $0.8-1.0$~GeV where it is generally agreed that thermal dileptons shine above hadronic and heavy quark decays (see e.g. Ref.~\cite{Rapp:2013nxa,Vujanovic:2013jpa}). At higher invariant mass, there are also wide regions where heavy quark decays are dominant, and can be used to study the plasma through heavy quark energy loss.

The expected importance of thermal emission is the reason behind the efforts being made to compute them with increasing accuracy. While these efforts have produced good agreement with dilepton measurements at the RHIC (e.g. Ref.~\cite{Vujanovic:2013jpa}), comparisons of thermal photon calculations with data are still not fully satisfactory. Recent calculations of thermal photons based on a hydrodynamical description of the plasma~\cite{Paquet:2015lta} underestimate the central values of the measured spectra by up to a factor of three at the RHIC, with less tension observed at the LHC. Keeping in mind the significant experimental uncertainties, this translate into a deviation of approximately one sigma or less at the LHC, and two sigma's or less at RHIC.  It must be emphasized that this tension is quantitative but not qualitative: the overall shape of the measured spectra is  described well by current thermal photon calculations.
As for direct photon $v_n$ measurements, there is tension with calculations at the RHIC, especially at larger $p_T^\gamma$, but also good agreement at lower $p_T^\gamma$. Moreover, the tension at higher $p_T^\gamma$ is understood to originate from the suppression of the thermal photon $v_2$ due to prompt photons (see e.g. Ref.~\cite{Paquet:2015lta} for more details). Calculations of the \emph{thermal} photon $v_2$ are actually very similar to direct photon $v_2$ measurements. Consequently, as far as thermal photons are concerned, the current tension with both spectra and $v_n$ measurements originate mainly from one cause: too few thermal photons are predicted by calculations.

In view of the above, it is still reasonable at the moment to work under the assumption that low $p_T^\gamma$ direct photons are dominated by thermal photons, despite the current tension with up-to-date calculations. Various proposals that have been made to address this situation are reviewed in the next sections.


\section{Thermal emission \& spacetime evolution}


Evaluating electromagnetic emission in heavy ion collisions requires a detailed spacetime description of the plasma, a description typically provided by hydrodynamics, although other approaches have been used~\cite{Linnyk:2015rco}. Hydrodynamical models can be summarized by an initial condition,
the equations of motion of relativistic viscous hydrodynamics, an equation of state,
transport coefficients such as shear and bulk viscosities, and a criteria to decide when to stop describing the plasma as its temperature decreases.
The usual assumption is that hydrodynamics provides a reasonable description of the plasma's expansion until its temperature drops below confinement, at which point hadronic degrees of freedom can be used. 
Hadronic transport models can then further describe the interactions of the resulting hadrons, and hadronic observables are evaluated once all interactions cease. 

\begin{figure}[tb]
	\centering
	\includegraphics[width=0.7\columnwidth]{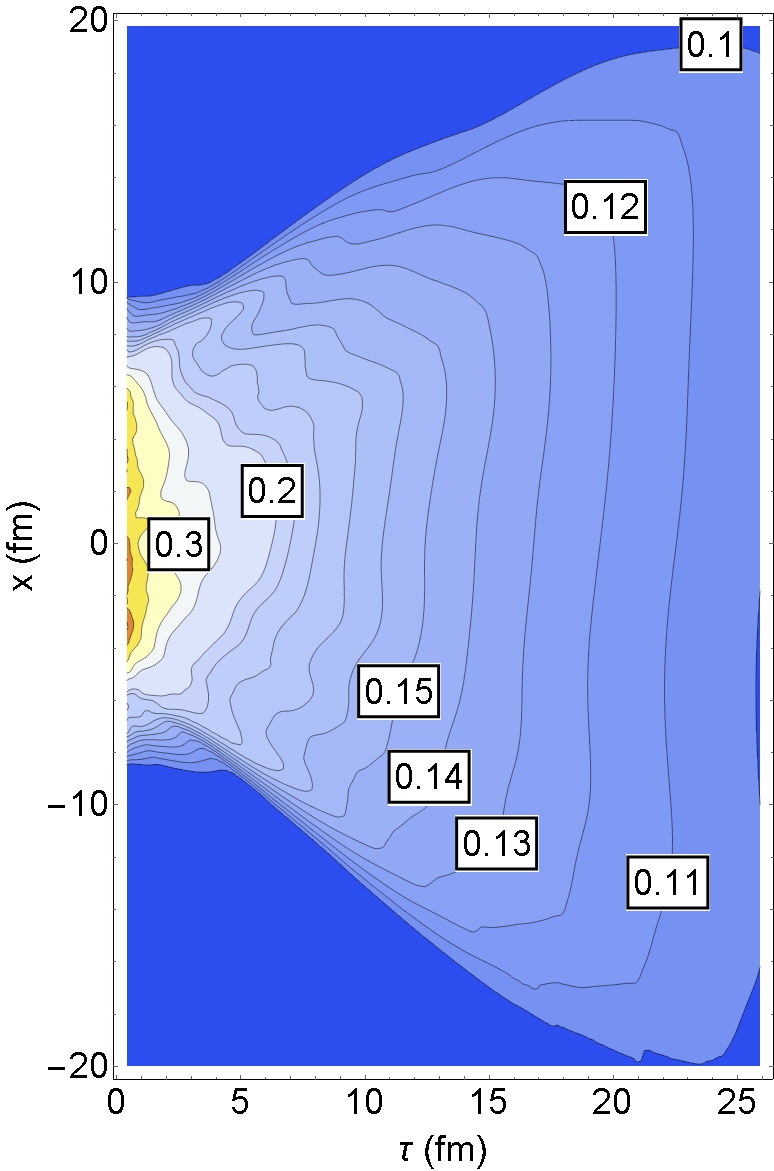}
	\caption{Temperature profile predicted by a hydrodynamic simulation for a single central Pb-Pb collision with $\sqrt{s_{NN}}=2760$~GeV. Contours are shown every 10 MeV for temperatures below 200 MeV, and every 50 MeV above. Labels shown for various contours are in GeV.}
	\label{fig:Tprofile}
\end{figure}

\begin{figure}[tb]
	\centering
	\includegraphics[width=0.7\columnwidth]{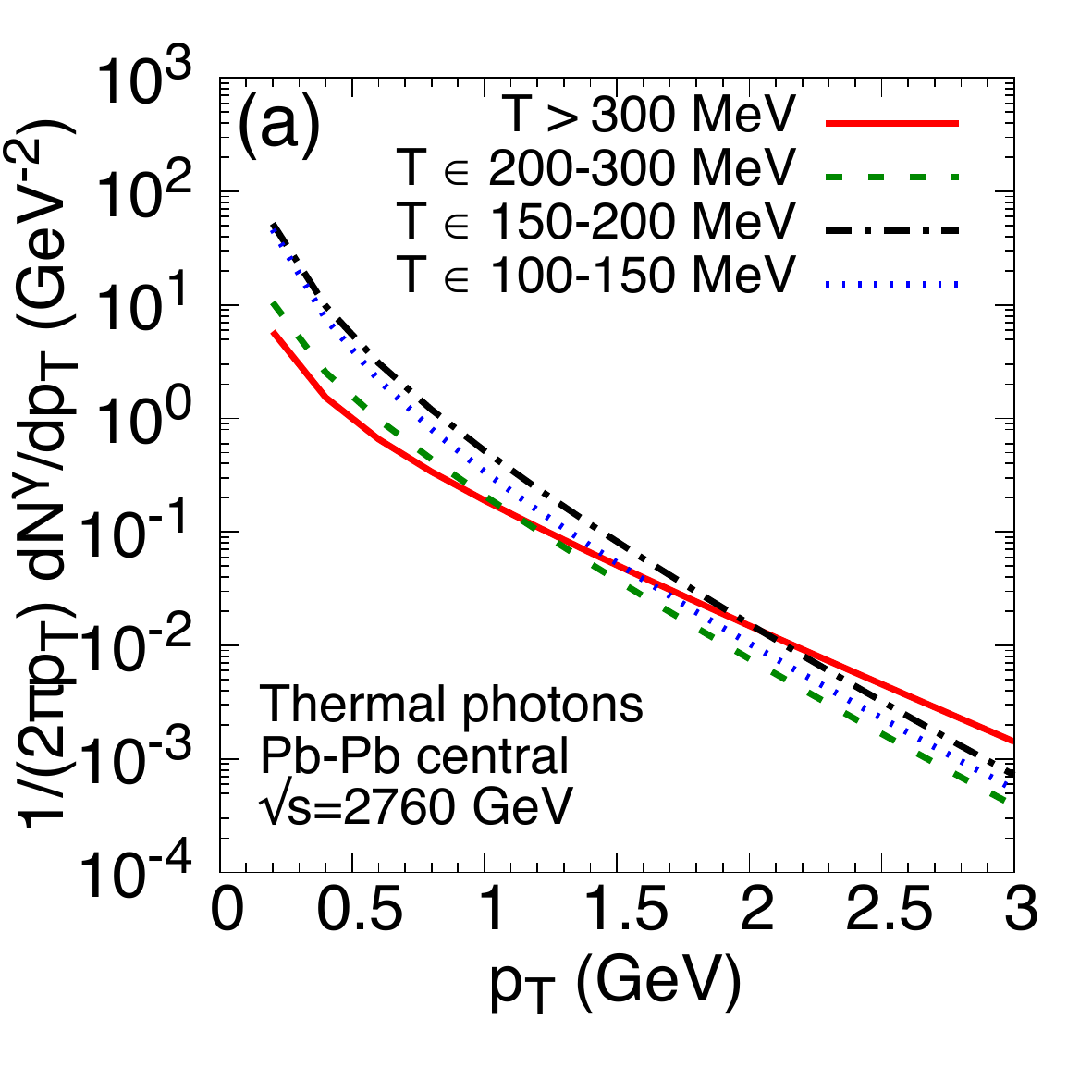}
	\includegraphics[width=0.7\columnwidth]{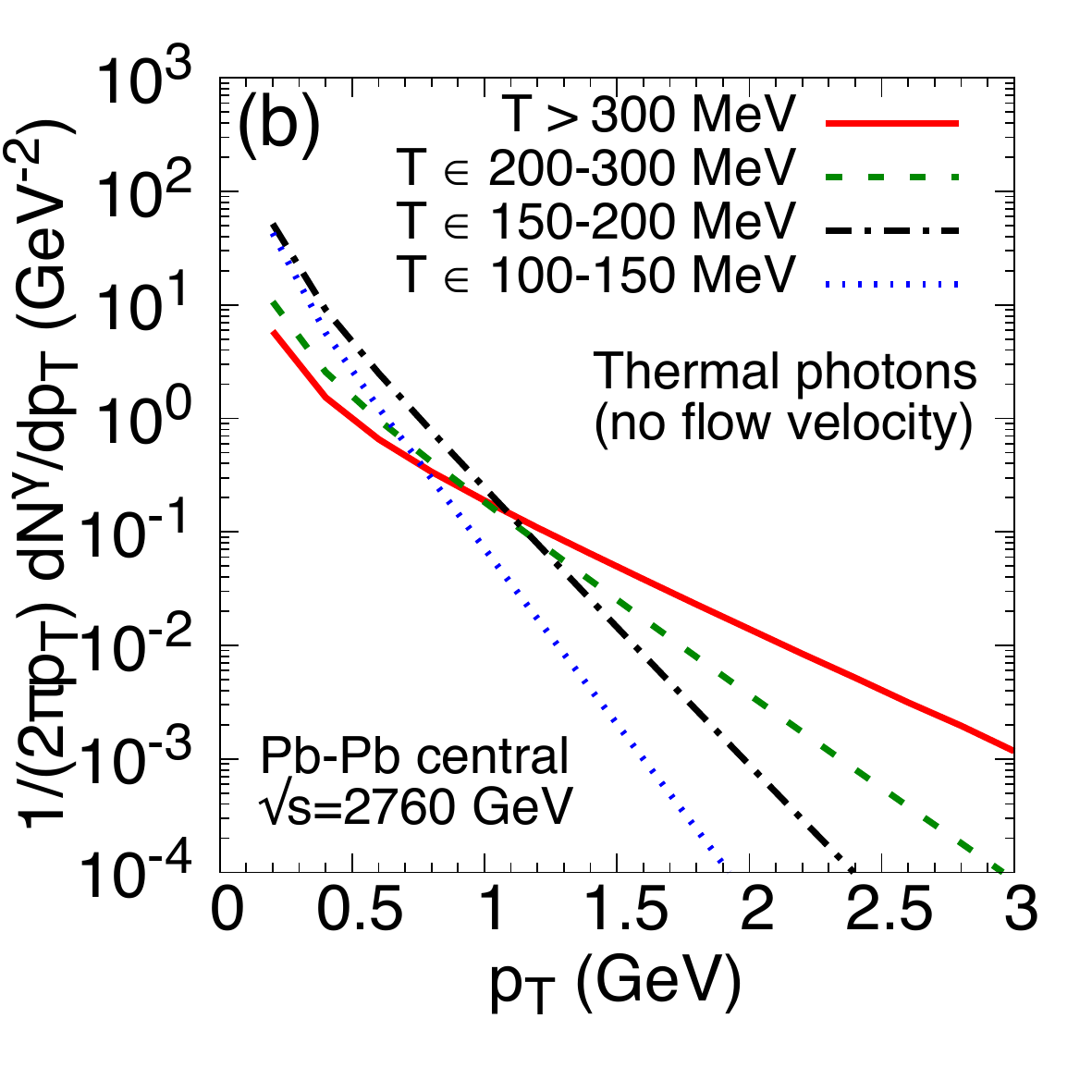}
	\caption{(a) Spectra of thermal photons produced in different range of temperatures, for the hydrodynamics event corresponding to Fig.~\ref{fig:Tprofile}. (b) Same as (a) without the effect of the flow velocity on photon emission. See text for details.}
	\label{fig:spectra}
\end{figure}

Initial conditions for hydrodynamics, usually provided at a fixed time $\tau=\sqrt{t^2-z^2}$, are either constrained by simplified first-principle descriptions of the early-time dynamics of heavy-ion collisions, or an ansatz like the Glauber model, in both case with some parameters to be adjusted to measurements. The ansatz approach can be sufficient to study many hadronic observables, which often have a limited sensitivity to smaller features of the initial conditions. This is not necessarily sufficient for electromagnetic probes, which are emitted at all times.
In this sense, a smooth transition from early time degrees of freedom to hydrodynamics is a desirable feature of hydrodynamics models of heavy ion collisions.

This continuous transition between the early plasma, the intermediate hydrodynamical expansion and the late time hadronic transport blurs the distinction between thermal emission and radiation produced when the medium is not necessarily very close to local equilibrium. 
To better illustrate this issue, the simulated temperature profile for a single central Pb-Pb collision ($\sqrt{s_{NN}}=2760$~GeV), as obtained with the hydrodynamical model used in Ref.~\cite{Paquet:2015lta,Ryu:2015vwa}, is shown in Fig.~\ref{fig:Tprofile} as a function of time and transverse direction. As is usually the case with hydrodynamic simulations, hadronic observables were used to constrain all parameters of the model. The entire spacetime evolution of the plasma is fixed and electromagnetic observables are computed \emph{a posteriori}. For this central event, the hottest regions of the plasma reach temperatures of 500-600~MeV at the point when the hydrodynamics simulation is started, $\tau_0=0.4$~fm. Figure~\ref{fig:Tprofile} shows how rapidly the plasma cools down as it expands. 


The production of thermal photons and dileptons is given by\footnote{Viscous corrections to the thermal rate are neglected in Eq.~\ref{eq:yield}, since they are generally modest on the thermal photon spectra, unlike for the $v_n$'s (see e.g. Ref.~\cite{Paquet:2015lta}). If included, $d^4 \Gamma_{\gamma/l^+l^-}/d^4 k$ would also depend on the bulk pressure $\Pi(X)$ and the shear tensor $\pi^{\mu\nu}(X)$.}
\begin{equation}
 \frac{d^4 N_{\gamma/l^+l^-}}{d^4 k}=\int d^4X \frac{d^4 \Gamma_{\gamma/l^+l^-}}{d^4 k}(K^\mu,u^\mu(X),T(X))
\label{eq:yield}
\end{equation}
where $T(X)$ is the temperature of the plasma, $u^\mu(X)$ the flow velocity, and the integral runs over the spacetime volume of the plasma.
%
%
For the event shown in Fig.~\ref{fig:Tprofile}, the spectra of thermal photons produced in different ranges of temperature is shown in Fig.~\ref{fig:spectra}(a). There are three factors
that impact thermal emission: the temperature, the spacetime volume and the flow velocity distribution. The temperature and spacetime volume distribution can be seen in Fig.~\ref{fig:Tprofile}. The flow velocity, not shown in Fig.~\ref{fig:Tprofile}, is small at early times but increases as the plasma expands. This flow has a large effect on thermal photon production, illustrated in Fig.~\ref{fig:spectra}(b) by showing what thermal production would be like without the effect of the flow velocity. Note that this is for illustration purpose only, and that it obviously does not represent a physical situation.

Figures~\ref{fig:Tprofile} and~\ref{fig:spectra} show that despite the very small spacetime volume with a temperature above 300~MeV, a significant number of photons are emitted, because of the large thermal photon emission rate $d^4 \Gamma_{\gamma}/d^4 k$ at high temperature. The effect of the  flow velocity for this range of temperatures is very small, as seen by the small difference between Fig.~\ref{fig:spectra}(a) and (b) (solid curve). 

The photon emission rate decreases with the temperature, but since the spacetime volume increases,   photon production remains high. Figure~\ref{fig:spectra}(b) shows how lower temperatures translate into steeper slopes for the photon spectra, as a result of the $\sim \exp(-p_T/T)$ dependence of the thermal rate $d^4 \Gamma_{\gamma}/d^4 k$. On the other hand, Fig.~\ref{fig:spectra}(a) shows how the flow velocity developed during the expansion significantly shift the higher $p_T^\gamma$ spectra for low temperature emissions, giving photons emitted in all temperature ranges a much more similar $p_T^\gamma$ slope. A detailed investigation of the effect of flow velocities on thermal photon production can be found in Refs~\cite{vanHees:2011vb,Shen:2013vja}.

Figures~\ref{fig:Tprofile} and ~\ref{fig:spectra} can be used to better illustrate the importance of photon emission at early and late times, which is discussed in the next sections.

\subsection{Late time electromagnetic emission}

Figures~\ref{fig:Tprofile} and ~\ref{fig:spectra} show that, if the plasma can maintain near local equilibrium at temperatures below confinement, significant electromagnetic emission is produced owing to the large volume and flow velocity predicted by hydrodynamics. It is however more common, and likely more appropriate, to describe the late stage of heavy ion collisions with hadronic degrees of freedom whose interactions are simulated with a transport model. Studies of photon production in heavy ion collisions with such a hybrid approach include Refs~\cite{Baeuchle:2009ep,Bauchle:2010ym} (see also references to previous works therein). More recent efforts include Ref. \cite{Staudenmaier:2016hmh}. Mixed approaches in which a transport model description is coarse-grained to obtain a near-thermal medium from which electromagnetic emission is evaluated have also been studied, see e.g. Refs~\cite{Huovinen:2002im,Endres:2016tkg}.

Looking at the effect of a late stage hadronic transport on both hadrons and electromagnetic emissions simultaneously is of prime importance: it is needed to clarify how much photons and dileptons are produced at low temperatures, and also how smooth a transition between hydrodynamics and transport can be obtained. Moreover it offers an opportunity to make connections between hydrodynamics and different descriptions of the plasma~\cite{Linnyk:2015rco}, with regard to both hadrons and electromagnetic probes.


\subsection{Early time dynamics \& electromagnetic emission}

Understanding the early time dynamics of the plasma and its thermalization and transition to hydrodynamics is a very active topic of research. While significant progress has been made in recent years (see Refs~\cite{Lappi:2015jka,Kurkela:2016vts,Gelis:2016upa}), obtaining a realistic first-principle description of the early plasma remains a challenge.

The hydrodynamic evolution shown in Fig.~\ref{fig:Tprofile} begins at time $\tau=0.4$~fm; there are no contributions in Fig.~\ref{fig:spectra} from photons emitted at earlier times. 
Most thermal photon and dilepton calculations in the literature include only emission past the time $\tau_0$ at which the hydrodynamic evolution begins\footnote{While emissions before time $\tau_0$ are not usually included in the literature, the value of $\tau_0$ does change from calculation to calculation. The temperature at which the hydrodynamic evolution is stopped also varies. This reflects in part the uncertainty in the current understanding of the early and late stages of heavy ion collisions.
}. This affects calculations of electromagnetic emissions on different levels.

Thermalisation is thought to be a gradual process, which implies that there must be an extended spacetime region $\mathcal{I}$ at early times where the medium is not too far from local equilibrium, and different hypersurfaces in $\mathcal{I}$ could serve as initial conditions for hydrodynamics. For simplicity, $\mathcal{I}$ can be approximated by a range of times: $\mathcal{I} \approx [\tau^{eq}_{lower},\tau^{eq}_{upper}]$. 
The width of this time window is not fully clear, and the starting time $\tau_0$ of hydrodynamics could reasonably be any value in this $[\tau^{eq}_{lower},\tau^{eq}_{upper}]$ range. By using $\tau_0 \approx \tau^{eq}_{lower}$, calculations could in theory include all thermal radiation produced while the plasma can reasonably be described with hydrodynamics, but $\tau^{eq}_{lower}$ is not known precisely. It is thus generally not clear how much near-thermal emission is missing in different calculations. However, as seen in Fig.~\ref{fig:Tprofile}(a), it is possible to say that early time emission is mostly at high $p_T^\gamma$ (above $1.5-2$~GeV) for photons, and is likely to be a modest contribution at lower values of $p_T^\gamma$. A more detailed investigation, with similar conclusions, can be found in Ref.~\cite{Chatterjee:2012dn}.

At times before $\tau^{eq}_{lower}$, the plasma is increasingly out of equilibrium, and if a temperature is nevertheless defined by coarse-graining, it should grow much more slowly than the approximate $T\propto \tau^{-1/3}$ dependence of hydrodynamics. Electromagnetic emission in this region must be evaluated in a different framework, and one should be careful with estimates based on thermal emission rates and spacetime volume arguments: electromagnetic emission at early times may not be dominated by soft (near-thermal) modes, but rather by harder modes which carry a significant fraction of the plasma's energy.


The properties of the early plasma after hydrodynamics becomes acceptable ($\tau\gtrsim \tau^{eq}_{lower}$ in the notation used above) are also important. 
The very earliest times are understood to be dominated by gluons. If the plasma reaches thermal equilibrium before chemical equilibrium, hydrodynamics could be applicable at early times although the plasma is still dominated by gluons. This would lead to an overall suppression of photons and dileptons emitted.
Recent investigations of this effect include Refs~\cite{Vovchenko:2016ijt, Monnai:2014kqa}. In the case of photons, this once again affects mainly higher $p_T^\gamma$ emissions.

To summarize the effect on photons, it can be said with reasonable certainty that early time thermal or near-thermal emission contribute mainly at higher $p_T^\gamma$ ($\gtrsim 1.5$--$2$~GeV). Questions of chemical equilibration of the early plasma, or photon emissions at times earlier than the usual initialization time $\tau_0$ of hydrodynamics, will probably be restricted to this upper range of $p_T^\gamma$. It is thus unlikely that they would explain a tension with measurements spread across a large range of $p_T^\gamma$. Moreover, such photons tend to have a much smaller $v_n$ then suggested by current measurements. Non-thermal photon emission at early time have not yet been studied very much for heavy ion collisions, and could in theory contribute at all $p_T^\gamma$. It is a question that will deserve further investigations. 




\section{Other developments}

The thermal electromagnetic emission rate, denoted $d^4 \Gamma_{\gamma}/d^4 k$ in Eq.~\ref{eq:yield}, is central in the evaluation of thermal radiation.
Evaluating this rate in the range of temperatures relevant for heavy ion collisions ($T\sim 100$--$600$~MeV) remains a theoretical challenge. For temperatures above confinement, recent efforts include the perturbative evaluation at next-to-leading order in $\alpha_s$ of the photon and dilepton rates~\cite{Ghiglieri:2013gia,Ghiglieri:2014kma,Ghisoiu:2014mha,Ghiglieri:2015nba}, comparisons of the perturbative photon rate with lattice calculations~\cite{Ghiglieri:2016tvj}, and increasing efforts to constrain the thermal rates with holographic approaches~\cite{Iatrakis:2016ugz,Finazzo:2015xwa}. Efforts below confinement include the calculation of additional hadronic emission channels~\cite{Holt:2015cda}. Progress has also been made in evaluating the corrections to the thermal rates due to shear and bulk viscosities~\cite{ViscousRate}.

As is clear from Fig.~\ref{fig:spectra}(a), thermal emissions at temperatures around and slightly above deconfinement are very large contributions to the total thermal signal. There is on-going debates on the magnitude of the thermal rates in this temperature range. Above confinement, some investigations suggest that the perturbative thermal photon rate is considerably too large~\cite{Finazzo:2015xwa,Gale:2014dfa}, or possibly too large only for low energy photons~\cite{Ghiglieri:2016tvj}, or  too large for low energy photons and too small for higher energy ones~\cite{Iatrakis:2016ugz}. It will be important to see if a common ground can be found, and also to better understand how much and in what range of temperatures holographic calculations can be relied on.

Prompt photons, produced in hard parton collisions at the initial nuclei impact, are certainly a dominant source of direct photons for $p_T^\gamma \gtrsim 3$~GeV, and possibly a non-negligible source at lower $p_T^\gamma$ as well. Photons from parton interactions with the plasma (``jet-medium photons''), which are closely related to the fragmentation photon component of prompt photons, most likely contribute to low $p_T^\gamma$ direct photons (see e.g. Ref.~\cite{Turbide:2007mi}). There has not been any recent attempts to evaluate simultaneously and consistently prompt and jet-plasma photons using the latest advances in parton energy loss and hydrodynamical modelling of heavy ion collisions. The implementation of prompt photon production at next-to-leading order in a Monte Carlo model~\cite{Jezo:2016ypn} could provide impetus for new studies of this important question.

Besides heavy ion collisions, there is also increasing interest in electromagnetic probes in smaller collision systems~\cite{WhyTheHellMciteIsntWorking,MciteIsStillNotWorking} and at lower collision energies such as those provided by the RHIC beam energy scan. These systems will provide a different angle of approach to study electromagnetic probes, and may help better understand collectivity in nuclear collisions.

\section{Summary and outlook}

Electromagnetic probes offer additional sensitivity to the properties of plasma across a broad range of temperatures: the early and hot phases of the plasma, the confinement region and possibly also the low temperature, late stage of the medium. Considerable progress has been made over the past years in the understanding of thermal electromagnetic probes. Future work on prompt and jet-medium photons, thermal emission rates around and above confinement, and early and late stage emission, will bring our understanding of electromagnetic probes to the next level, and shed additional light on tensions observed with measurements. Time is also ripe for simultaneous studies of photon and dilepton production using recent developments in our understanding of heavy ion collisions, to clarify what is the status of these closely related observables.

Should increasingly accurate measurements of direct photons and refined calculations continue to show tension, new interesting possibilities will have to be considered more closely. This include the contribution of more exotic sources of electromagnetic probes, and modifications to the current picture of the evolution of heavy ion collisions. Electromagnetic probes promise plenty of opportunities to push the boundaries of our current understanding of heavy ion collisions and QCD.

\textbf{Acknowledgements} 

I thank Charles Gale, S\"oren Schlichting, Chun Shen, Edward Shuryak, Ismail Zahed and the participants of Hard Probes 2016 for interesting discussions and feedback. This work was supported by the U.S. D.O.E. Office of Science, under Award No. DE-FG02-88ER40388, and by a Stony Brook University Office of Postdoctoral Affairs travel grant. 




\bibliographystyle{elsarticle-num}
\bibliography{jos}

\begin{thebibliography}{10}
\expandafter\ifx\csname url\endcsname\relax
  \def\url#1{\texttt{#1}}\fi
\expandafter\ifx\csname urlprefix\endcsname\relax\def\urlprefix{URL }\fi
\expandafter\ifx\csname href\endcsname\relax
  \def\href#1#2{#2} \def\path#1{#1}\fi

\bibitem{deSouza:2015ena}
R.~Derradi~de Souza, T.~Koide, T.~Kodama, {Hydrodynamic Approaches in
  Relativistic Heavy Ion Reactions}, Prog. Part. Nucl. Phys. 86 (2016) 35--85.

\bibitem{Gale:2009gc}
C.~Gale, {Photon Production in Hot and Dense Strongly Interacting Matter},
  Landolt-Bornstein 23 (2010) 445.

\bibitem{preEq}
{See contributions of M. Greif and M. Ruggieri in these proceedings.}

\bibitem{BatheHP2016}
{See contribution of S. Bathe in these proceedings, and references therein, for
  an up-to-date overview of electromagnetic measurements at the RHIC and the
  LHC}.

\bibitem{Bock:2016jus}
F.~Bock, C.~Loizides, T.~Peitzmann, M.~Sas, {Impact of residual contamination
  on inclusive and direct photon flow. }\href {http://arxiv.org/abs/1606.06077}
  {\path{arXiv:1606.06077}}.

\bibitem{Turbide:2007mi}
S.~Turbide, C.~Gale, E.~Frodermann, U.~Heinz, {Electromagnetic radiation from
  nuclear collisions at RHIC energies}, Phys. Rev. C77 (2008) 024909.

\bibitem{Rapp:2013nxa}
R.~Rapp, {Dilepton Spectroscopy of QCD Matter at Collider Energies}, Adv. High
  Energy Phys. 2013 (2013) 148253.

\bibitem{Vujanovic:2013jpa}
G.~Vujanovic, C.~Young, B.~Schenke, R.~Rapp, S.~Jeon, C.~Gale, {Dilepton
  emission in high-energy heavy-ion collisions with viscous hydrodynamics},
  Phys. Rev. C89~(3) (2014) 034904.

\bibitem{Paquet:2015lta}
J.-F. Paquet, C.~Shen, G.~S. Denicol, M.~Luzum, B.~Schenke, S.~Jeon, C.~Gale,
  {Production of photons in relativistic heavy-ion collisions}, Phys. Rev.
  C93~(4) (2016) 044906.

\bibitem{Linnyk:2015rco}
O.~Linnyk, E.~L. Bratkovskaya, W.~Cassing, {Effective QCD and transport
  description of dilepton and photon production in heavy-ion collisions and
  elementary processes}, Prog. Part. Nucl. Phys. 87 (2016) 50--115.

\bibitem{Ryu:2015vwa}
S.~Ryu, J.-F. Paquet, C.~Shen, G.~S. Denicol, B.~Schenke, S.~Jeon, C.~Gale,
  {Importance of the Bulk Viscosity of QCD in Ultrarelativistic Heavy-Ion
  Collisions}, Phys. Rev. Lett. 115~(13) (2015) 132301.

\bibitem{vanHees:2011vb}
H.~van Hees, C.~Gale, R.~Rapp, {Thermal Photons and Collective Flow at the
  Relativistic Heavy-Ion Collider}, Phys. Rev. C84 (2011) 054906.

\bibitem{Shen:2013vja}
C.~Shen, U.~W. Heinz, J.-F. Paquet, C.~Gale, {Thermal photons as a quark-gluon
  plasma thermometer reexamined}, Phys. Rev. C89~(4) (2014) 044910.

\bibitem{Baeuchle:2009ep}
B.~{B\"auchle}, M.~Bleicher, {Hybrid model calculations of direct photons in
  high-energy nuclear collisions}, Phys. Rev. C81 (2010) 044904.

\bibitem{Bauchle:2010ym}
B.~{B\"auchle}, M.~Bleicher, {Direct photon calculations in heavy-ion
  collisions at $\sqrt{s_{NN}}$ = 62.4 - 200 AGeV in a (3+1) dimensional hybrid
  approach}, Phys. Rev. C82 (2010) 064901.

\bibitem{Staudenmaier:2016hmh}
J.~Staudenmaier, J.~Weil, H.~Petersen, {Non-equilibrium dilepton production in
  hadronic transport approaches. }\href {http://arxiv.org/abs/1611.09164}
  {\path{arXiv:1611.09164}}.

\bibitem{Huovinen:2002im}
P.~Huovinen, M.~Belkacem, P.~J. Ellis, J.~I. Kapusta, {Dileptons and photons
  from coarse grained microscopic dynamics and hydrodynamics compared to
  experimental data}, Phys. Rev. C66 (2002) 014903.

\bibitem{Endres:2016tkg}
S.~Endres, H.~van Hees, M.~Bleicher, {Energy, centrality and momentum
  dependence of dielectron production at collider energies in a coarse-grained
  transport approach}, Phys. Rev. C94~(2) (2016) 024912.

\bibitem{Lappi:2015jka}
T.~Lappi, Initial state in heavy ion collisions, Nuclear and Particle Physics
  Proceedings 276–278 (2016) 29 -- 34.

\bibitem{Kurkela:2016vts}
A.~Kurkela, {Initial state of Heavy-Ion Collisions: Isotropization and
  thermalization}, Nucl. Phys. A956 (2016) 136--143.

\bibitem{Gelis:2016upa}
F.~Gelis, B.~Schenke, Initial-state quantum fluctuations in the little bang,
  Annual Review of Nuclear and Particle Science 66~(1) (2016) 73--94.

\bibitem{Chatterjee:2012dn}
R.~Chatterjee, H.~Holopainen, T.~Renk, K.~J. Eskola, {Collision centrality and
  $\tau_0$ dependence of the emission of thermal photons from fluctuating
  initial state in ideal hydrodynamic calculation}, Phys. Rev. C85 (2012)
  064910.

\bibitem{Vovchenko:2016ijt}
V.~Vovchenko, I.~A. Karpenko, M.~I. Gorenstein, L.~M. Satarov, I.~N. Mishustin,
  B.~{K\"ampfer}, H.~Stoecker, {Electromagnetic probes of a pure-glue initial
  state in nucleus-nucleus collisions at energies available at the CERN Large
  Hadron Collider}, Phys. Rev. C94~(2) (2016) 024906.

\bibitem{Monnai:2014kqa}
A.~Monnai, {Thermal photon $v_2$ with slow quark chemical equilibration}, Phys.
  Rev. C90~(2) (2014) 021901.

\bibitem{Ghiglieri:2013gia}
J.~Ghiglieri, J.~Hong, A.~Kurkela, E.~Lu, G.~D. Moore, D.~Teaney,
  {Next-to-leading order thermal photon production in a weakly coupled
  quark-gluon plasma}, JHEP 05 (2013) 010.

\bibitem{Ghiglieri:2014kma}
J.~Ghiglieri, G.~D. Moore, {Low Mass Thermal Dilepton Production at NLO in a
  Weakly Coupled Quark-Gluon Plasma}, JHEP 12 (2014) 029.

\bibitem{Ghisoiu:2014mha}
I.~Ghisoiu, M.~Laine, {Interpolation of hard and soft dilepton rates}, JHEP 10
  (2014) 83.

\bibitem{Ghiglieri:2015nba}
J.~Ghiglieri, {The thermal dilepton rate at NLO at small and large invariant
  mass}, Nucl. Part. Phys. Proc. 276-278 (2016) 305--308.

\bibitem{Ghiglieri:2016tvj}
J.~Ghiglieri, O.~Kaczmarek, M.~Laine, F.~Meyer, {Lattice constraints on the
  thermal photon rate}, Phys. Rev. D94~(1) (2016) 016005.

\bibitem{Iatrakis:2016ugz}
I.~Iatrakis, E.~Kiritsis, C.~Shen, D.-L. Yang, {Holographic Photon Production
  in Heavy Ion Collisions. }\href {http://arxiv.org/abs/1609.07208}
  {\path{arXiv:1609.07208}}.

\bibitem{Finazzo:2015xwa}
S.~I. Finazzo, R.~Rougemont, {Thermal photon, dilepton production, and electric
  charge transport in a baryon rich strongly coupled QGP from holography},
  Phys. Rev. D93~(3) (2016) 034017.

\bibitem{Holt:2015cda}
N.~P.~M. Holt, P.~M. Hohler, R.~Rapp, {Thermal photon emission from the πρω
  system}, Nucl. Phys. A945 (2016) 1--20.

\bibitem{ViscousRate}
{See contributions of S. Hauksson and G. Vujanovic in these proceedings.}

\bibitem{Gale:2014dfa}
C.~Gale, Y.~Hidaka, S.~Jeon, S.~Lin, J.~Paquet, R.~D. Pisarski, D.~Satow, V.~V.
  Skokov, G.~Vujanovic, {Production and Elliptic Flow of Dileptons and Photons
  in a Matrix Model of the Quark-Gluon Plasma}, Phys. Rev. Lett. 114 (2015)
  072301.

\bibitem{Jezo:2016ypn}
T.~Jezo, M.~Klasen, F.~{K\"onig}, {Prompt photon production and photon-hadron
  jet correlations with POWHEG}, JHEP 11 (2016) 033.

\bibitem{WhyTheHellMciteIsntWorking}
{C. Shen, J. F. Paquet, G. S. Denicol, S. Jeon, C. Gale, Thermal photon
  radiation in high multiplicity p+Pb collisions at the Large Hadron Collider,
  Phys. Rev. Lett. 116 (7) (2016) 072301; C. Shen, J.-F. Paquet, G. S. Denicol,
  S. Jeon, C. Gale, Collectivity and electromagnetic radiation in small
  systems. arXiv:1609.02590; See also the contribution of C. Shen in these
  proceedings.}

\bibitem{MciteIsStillNotWorking}
{S. Beni\'c, K. Fukushima, O. Garcia-Montero, R. Venugopalan, Probing gluon
  saturation with next-to-leading order photon production at central rapidities
  in proton-nucleus collisions. arXiv:1609.09424; See also the contribution of
  S. Beni\'c in these proceedings.}

\end{thebibliography}







\end{document}